\documentclass[prb,aps,onecolumn,showpacs,amsmath,amssymb,print,unsortedaddress,floatfix]{revtex4-1}
\usepackage{graphicx}
\usepackage{epstopdf}
\usepackage{tabularx}
\usepackage{color}
\usepackage{natbib}

\newcommand{\Ang}{\mathrm{\AA}}

\def\eV{\,\textrm{eV}}

\newcommand{\hbn}{{\it h}-BN}

\def\Im{\mathop{\textrm{Im}}}

\begin{document}

\title{
Electronic resistances of multilayered two-dimensional crystal junctions
}

\author{Yun-Peng Wang,$^{1,2} $ X.-G. Zhang,$^{1,2}$  J. N. Fry,$^1$  and Hai-Ping Cheng$^{1,2}$ } 
\email[Corr. author: Hai-Ping Cheng,  ]{ hping@ufl.edu }
\affiliation{$^1$Department of Physics, University of Florida,  Gainesville, Florida 32611, USA \\
             $^2$Quantum Theory Project, University of Florida, Gainesville, Florida 32611, USA}

\begin{abstract}

We carry out a layer-by-layer investigation to understand electron transport across metal-insulator-metal junctions.
Interfacial structures of junctions were studied and characterized using first-principles density functional theory 
within the generalized gradient approximation.
We found that as a function of the number of crystal layers 
the calculated transmission coefficients of multilayer silicene junctions decay much slower 
than for \hbn-based junctions 
We revisited the semiclassical Boltzmann theory of electronic transport and applied to multilayer silicene and \hbn-based junctions.
The calculated resistance in the high-transmission regime is smaller than that provided by the Landauer formula.
As the thickness of the barrier increases, results from the Boltzmann and the Landauer formulae converge.
We provide a upper limit in the transmission coefficient below which, the Landauer method becomes valid.
Quantitatively, when the transmission coefficient is lower than $ \sim 0.05 $ per channel,
the error introduced by the Landauer formula for calculating the resistance is negligible.
In addition, we found that the resistance of a junction is not entirely determined by the averaged transmission,
but also by the distribution of the transmission over the first Brillouin zone.

\end{abstract}

\maketitle

\section{introduction}

Silicene is a single atomic layer of Si atoms arranged in a two-dimensional buckled honeycomb lattice.
Although it was predicted decades ago that free-standing silicene exhibits a massless relativistic behavior near the Fermi energy,
\cite{PhysRevB.50.14916, PhysRevB.72.075420,PhysRevB.76.075131, PhysRevLett.102.236804}
it was only very recently that silicene was synthesized on the surfaces of a few metallic substrates.%
\cite{PhysRevLett.108.155501,PhysRevLett.109.056804,PhysRevLett.108.245501,NanoLett.13.685}
Linear electronic dispersions observed in the silicene/Ag(111) system were attributed 
as the signature of Dirac fermions in silicene.\cite{PhysRevLett.108.155501,PhysRevLett.109.056804}
Subsequent experimental and theoretical studies revealed the absence of Dirac fermions near the Fermi energy; 
instead, the observed linear electronic dispersions are  $sp$-bands of Ag or hybrid interface states.%
\cite{PhysRevLett.110.076801,PhysRevB.87.235435,JPSJ.82.063714,PhysRevB.87.245430,JApplPhys.114.113710,ApplPhysLett.103.231604}

Currently there is a growing interest in multilayer silicenes.%
\cite{ApplPhysLett.102.163106,ApplPhysLett.104.021602,JPCM.25.382202,JPCM.26.185003,JPCM.25.085508,PhysRevB.89.155418,PhysRevB.89.241403}
The atomic structure of multilayer-silicene has not yet been identified experimentally.
The surface morphology of multilayer-silicene on Ag(111) was examined
using scanning tunneling microscopy (STM) and scanning electron microscopy (SEM).
Growth of the first silicene layer on an Ag(111) substrate forms a $ 4 \times 4 $ super cell with respect to the Ag(111) surface
and $ 3 \times 3 $ with respect to free-standing silicene.
For multilayer silicene, a $ (\sqrt{3} \times \sqrt{3}\,) \mathrm{R}30^\circ$
reconstruction with respect to the free-standing silicene was observed.%
\cite{ApplPhysLett.102.163106,ApplPhysLett.104.021602,JPCM.25.382202,JPCM.26.185003,JPCM.25.085508,PhysRevB.89.155418,PhysRevB.89.241403}
The reconstruction in multilayer-silicene/Ag(111) was reproduced theoretically,
and it was predicted that only the surface silicene layer reconstructs.\cite{PhysRevB.89.155418}

The electronic transport properties of silicenes are important for their possible applications.
The current-in-plane (CIP) configuration is a natural choice to measure the resistance of silicenes, 
but the current tends to pass through the highly conducting Ag(111) substrates,
which makes measurements of the CIP resistances rather difficult.\cite{ApplPhysLett.104.021602}
Moreover, multilayer silicenes grow in the so-called Stranski-Krastanov mode (also known as the layer-plus-island mode), 
in which\cite{ApplPhysLett.102.163106,ApplPhysLett.104.021602}
the first silicene layer forms a continuous film on Ag(111),
while subsequent Si atoms form islands of multilayer-silicene.
As a result, it is difficult to directly associate the measured CIP resistance with the thickness of a multilayer silicene.
It is much easier to perform a current-perpendicular-to-plane (CPP) measurement
by placing a conducting probe on top of a multilayer-silicene island and measuring the resistance between the probe and the Ag substrate.

The transport properties of ultrathin silicene-based junctions were studied previously,\cite{PhysRevB.88.125428}
and the calculated average transmission per channel are about $ 0.7 $ and $ 0.35 $ 
for monolayer and bilayer silicene based junctions, respectively.
The Landauer formula is no longer accurate for calculating the resistances of junctions with such high transmissions, 
and we instead applied the semiclassical Boltzmann (SCB) theory to calculate the resistance.
In this work we extended our previous work to study silicene-based junctions with up to eight silicene layers.
The comparison between calculated SCB and Landauer resistances reveals the relation between them,
and more importantly provides an empirical threshold for the application of the Landauer formula.

The rest of the paper is organized as follows.
The computation method is presented in Section~\ref{sec:comput}.
The calculation results for Ag(111)$|$multilayer-silicene$|$Ag junctions are shown in Section~\ref{sec:silicene}.
Complementary calculations on multilayer \hbn-based junctions are presented in Sec.~\ref{sec:hbn}.
A summary is given in Section~\ref{sec:summary}.

%%%%%%%%%%%%%%%%%%%%%%%%%%%%%%%%%%%%%
\section{computational method}
\label{sec:comput}
%%%%%%%%%%%%%%%%%%%%%%%%%%%%%%%%%%%%%

%The DFT complex band structures of multilayer silicenes were calculated using the \texttt{PWcond} program\cite{PhysRevB.70.045417}.
The atomic model for Ag(111)$|$multilayer-silicene$|$Ag(111) junctions was built  
according to the first-principles simulations of bilayer-silicene/Ag(111) interfaces 
presented in Ref.~\onlinecite{PhysRevB.89.155418} 
without considering atomic reconstructions at the interfaces.
The junctions with two to eight silicene layers were optimized 
using the projector-augmented wave\cite{PhysRevB.50.17953,PhysRevB.59.1758}  (PAW) 
based density functional theory (DFT) as implemented in the Vienna {\it ab initio} simulation package 
{\sc vasp}.\cite{CMS.6.15,PhysRevB.54.11169}
 In this work we employed the generalized gradient approximation (GGA) with the Perdew-Burke-Ernzerhof (PBE) 
parametrization\cite{PhysRevLett.77.3865}
The supercells used in the structural optimizations consist of the multilayer silicene and
five Ag(111) atomic layers with Ag atoms in the central atomic layer kept fixed in their bulk positions.
The lattice constant of free-standing monolayer-silicene is approximately $4/3$ times of the Ag(111) surface, 
and a supercell consisting of $4 \times 4$ Ag(111) layers and $ 3 \times 3 $ multilayer-silicene primitive unit cells
in the $x$-$y$ plane was used to simulate these junctions.\cite{PhysRevB.87.245430,PhysRevB.88.125428}
The lattice constant in the $x$-$y$ plane was fixed by that of Ag(111), 
 calculated using the PBE functional to be $ 2.952\,\Ang $; 
and the lattice constant of multilayer-silicene was $ 2.952\,\Ang \times 4/3 =  3.936\,\Ang $.
During structure optimizations, the height of the supercell (along the $z$-direction) and the coordinates of atoms were fully relaxed,
until the forces on unfixed atoms were smaller than $ 0.01\,\eV/\Ang $.

The DFT-based NEGF\cite{Datta1995,PhysRevB.63.245407,Chem.Phys.281.151} method
was then used to compute the Green's function and the total transmission of these junctions.
%The scattering region was chosen to include the multilayer silicenes and the three adjacent Ag(111) layers in each lead.
NEGF calculations were performed using the {\sc transiesta} code\cite{PhysRevB.65.165401}.
Numerical atomic orbitals were used to expand the Hamiltonian and the Green's function.
Single-zeta plus polarization orbital (SZP) and double-zeta plus polarization orbital basis sets (DZP) for Ag and Si respectively
were generated using the default parameters in {\sc siesta}.\cite{SIESTA}
Norm-conserving pseudopotentials\cite{PhysRevB.43.1993} were used to describe interactions
between valence electrons ($ {3s^2 3p^2} $ for Si and $ {4d^{10} 5s^1} $ for Ag) and the corresponding core electrons.
The direction of transport was chosen as the $z$-direction.
Translational symmetry in the $x$-$y$ plane was exploited by using
a $ 15 \times 15 $ $k$-point mesh for calculating the charge density and $ 55 \times 55 $ for the Green's function.

The group velocity of the Bloch states in the Ag(111) leads along the $z$-direction is\cite{PhysRevB.88.125428}
\begin{equation}
\label{eq:groupv}
v_z^j({\bf k}_\|) = \frac{a_z}{\hbar}
\left[ u^j({\bf k}_\|) \right]^{\dagger} \Gamma^{\mathrm{lead}} u^j({\bf k}_\|),
\end{equation}
where $a_z$ is the length of the unit cell of the Ag lead; 
$u^j({\bf k}_\|)$ is the periodic part of the Bloch waves;
${\bf k}_\|$ are the components of $k$-points in the $x$-$y$ plane; 
$j$ is the index for Bloch waves at the same ${\bf k}_\|$; and 
$\Gamma^{\mathrm{lead}} = i(\Sigma^r - \Sigma^a) $, where
$\Sigma^r$ and $\Sigma^a$ are the retarded and advanced self-energy of the Ag leads, respectively.
The transmission $t$ and reflection $r$ coefficients corresponding to the Bloch waves in the left and the right leads 
are\cite{PhysRevB.88.125428} 
\begin{equation}
\label{eq:transmission}
t^{M,N}_{j,j'} = \frac{i}{\hbar \sqrt{|v^{M,j}_{z,<}| \,  |v^{N,j'}_{z,>}|} }
\left( u^{N,j'}_{>} \right)^{\dagger} \, \Gamma^{\mathrm{lead}}_N
G^r_{NM} \Gamma^{\mathrm{lead}}_M (u^{M,j}_{<}),
\end{equation}
\begin{equation}
\label{eq:reflection}
r^{M,M}_{j,j'} = \frac{1}{\hbar \sqrt{|v^{M,j}_{z,<}| \,  |v^{M,j'}_{z,>}|} }
\left[ i ( u^{M,j'}_{>} )^{\dagger}  \, \Gamma^{\mathrm{lead}}_M
G^r_{MM} \Gamma^{\mathrm{lead}}_M (u^{M,j}_{<})
-  ( u^{M,j'}_{>}  )^{\dagger} \Gamma^{\mathrm{lead}}_M (u^{M,j}_{<}) \right].
\end{equation}
where the dependence on ${\bf k}_\|$ is omitted.
%
%{\color{blue}
The label $M$ denotes one of the Ag leads (left or right) and $N$ the other one; 
%}
the subscripts $>$ and $<$ denote Bloch waves propagating against and towards the junction, respectively; 
and $G^r_{MN}$ is the submatrix of the retarded Green function of the junctions.

The group velocities of Bloch waves in the leads [Eq.~(\ref{eq:groupv})] 
and the transmission and reflection coefficients [Eqs.~(\ref{eq:transmission}), (\ref{eq:reflection})] 
extracted from first-principles calculations are used as parameters for the Boltzmann equation.
%
%{\color{blue}
It is convenient to introduce an auxiliary quantity $h$ to characterize the change 
of the distribution function $f$  from its equilibrium $f_0$ with energy,
\begin{equation}
\label{eq:define_h}
f^j(z,{\bf k}_\|,E) = f_0(E) - h^j(z,{\bf k}_\|) \frac{\partial f_0}{\partial E}  \,.
\end{equation}
%} 
%
The Boltzmann equation for the CPP configuration\cite{JS.13.221, JAP.87.5173} is
\begin{equation}
\label{eq:boltzcpp}
\left[     v_{z}^j({\bf k}_{\|}) \frac{\partial}{\partial z}  + \frac{1}{\tau} \right]  h^j(z,{\bf k}_{\|})
 - \frac{\mu(z)}{\tau}
= -e v^j_{z}({\bf k}_{\|}) {\cal E}_z,
\end{equation}
where $\tau$ is the relaxation time in Ag leads; 
$\mu(z)$ and ${\cal E}_z$ denote the applied electric field and the chemical potential along the $z$-direction, respectively.
The method to solve the Boltzmann equation Eq.~(\ref{eq:boltzcpp}) numerically is given in the Appendix.

\begin{figure}[t]
\includegraphics[width=0.5\linewidth]{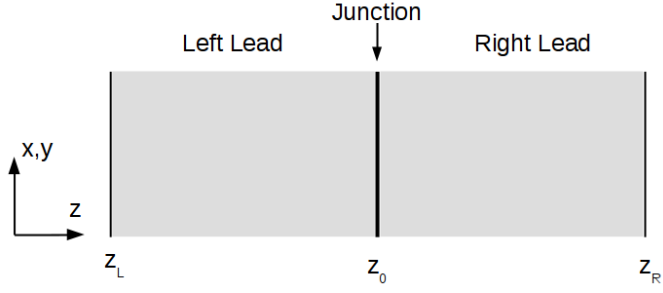}
\caption{
\label{fig:schematic}  
(Color online)
Schematics of a system in the current-perpendicular-to-plane (CPP) configuration.
The system consists of $ n $ metallic leads which are separated by $n-1$  junctions.
The system is homogeneous in the $x$-$y$ plane, and the electrical current flows in the $ z $ direction.
There are only two leads and one junction, {\it i.e.}, $n=2$
in our calculations on multilayer silicene and \hbn-based junctions.
}
\end{figure}

The current density along the $z$-direction is 
\begin{equation}
 {J} = -\frac{e}{2\pi \hbar A}
\sum_{{\bf k}_{\|},j}
\mathrm{sgn} \left[    {v^j_z}({\bf k_{\|}}) \right]
 h^j(z,{{\bf k}_{\|}})
 \equiv \sum_{{\bf k}_{\|}} J_{{\bf k}_{\|}}(z),
\label{eq:mcurrcpp}
\end{equation}
with $A$ the cross section of the unit cell of junction perpendicular to the $z$-direction.
%{\color{blue}
The total current density $J$ is a constant due to the conservation of charge, 
although each of its components $J_{{\bf k}_{\|}}(z)$ is not necessarily a constant.
%}
%
%{\color{blue}
The expression for the local chemical potential (see Appendix B) is
%} %% blue
$ \mu(z) =
\left\langle h^j(z,{\bf k}_\|) \right\rangle_{j,{\bf k}_\|}$,
and the voltage drop across the junction located at $z=z_0$ is equal to $\Delta V = \mu(z_0+0^+)-\mu(z_0-0^-)$;
thus the four-probe resistance of the junction is calculated as $\Delta V/J$.

%%%%%%%%%%%%%%%%%%%%%%%%%%%%%%%%%%%%%%%%%%%%%%%%%%
\section{multilayer silicene junctions}
%%%%%%%%%%%%%%%%%%%%%%%%%%%%%%%%%%%%%%%%%%%%%%%%%%
\label{sec:silicene}

\begin{figure}[h]
\includegraphics[width=0.2\linewidth]{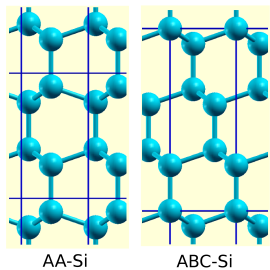}
\caption{
\label{fig:hex-fcc}  
(Color online)
Atomic structures of bulk AA-Si and ABC-Si. 
The unit cell boundary is denoted by blue lines.
}
\end{figure}

We considered two different stacking orders for the multilayer-silicenes.\cite{JPCM.25.085508}
In the first stacking order, two inequivalent silicene layers are stacked in an ``AA'' manner (denoted as ``AA-Si''),
and each Si atom can find another in its neighbouring layers with the same in-plane position.
The second stacking order, denoted as ``ABC-Si'', corresponds to the stacking along the (111) direction of diamond-structured silicon.
We note that both of these stacking configurations lead to a tetragonal arrangement of Si atoms,
as shown in Fig.~\ref{fig:hex-fcc}.

%\subsection{Energetics}

\begin{figure}[h]
\includegraphics[width=0.5\linewidth]{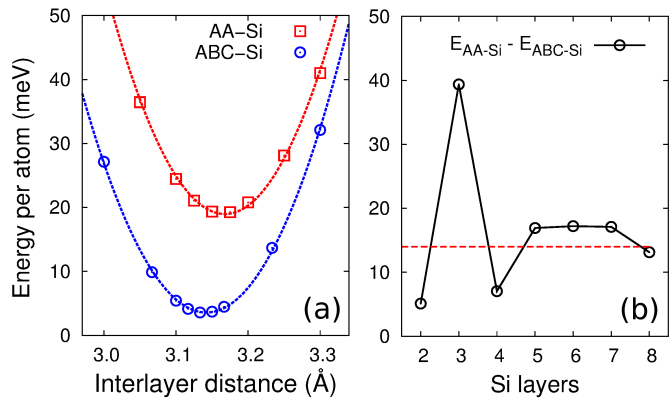}
\caption{
\label{fig:energy}  
(Color online)
Energetics of (a) bulk AA-Si and ABC-Si,
and (b) Ag(111)$|$AA-Si$|$Ag and Ag(111)$|$ABC-Si$|$Ag junctions as a function of the number of Si layers.
}
\end{figure}

The total energies of bulk AA- and ABC-Si were calculated as a function of the interlayer distance.
The interlayer distance is one half (third) of the lattice constant of bulk AA-Si (ABC-Si) along the $z$-direction.
Bulk AA-Si has a higher total energy than bulk ABC-Si by $ 14 \, \mathrm{meV} $ per Si atom, as shown in Fig.~\ref{fig:energy}(a).

We also calculated the total energies of multilayer silicene based junctions.
The total energies of AA-Si based junctions are always higher than the corresponding ABC-Si junctions,
and the total energy difference per Si atom is shown in Fig.~\ref{fig:energy}(b).
The total energy difference between bulk AA-Si and ABC-Si is denoted as the dashed line in Fig.~\ref{fig:energy}(b).
The deviations from the dashed line in Fig.~\ref{fig:energy}(b) are due to the interface effect.
We had also tried another method to obtain the atomic structures of junctions, in which 
the junction with ($N+1$) silicene layers was constructed and optimized 
from the junction with $N$ silicene layers by inserting a flat Si layer between the $N$th layer of silicene and the Ag(111) lead.
The resulting multilayer silicene  structures are different from either AA-Si or ABC-Si,
but are similar to the body-centered tetragonal $\mathrm{C_4}$ allotrope of carbon\cite{PhysRevLett.104.125504}.
Results of these structures are not shown due to their much higher energies.

%%%%%%%%%%%%%%%%%%%%%%%%%%%%%%%%%%%%%%%%%%%%%%%%%%%%%%
\subsection{Transmission}
%%%%%%%%%%%%%%%%%%%%%%%%%%%%%%%%%%%%%%%%%%%%%%%%%%%%%%
\begin{figure}[h]
\includegraphics[width=0.5\linewidth]{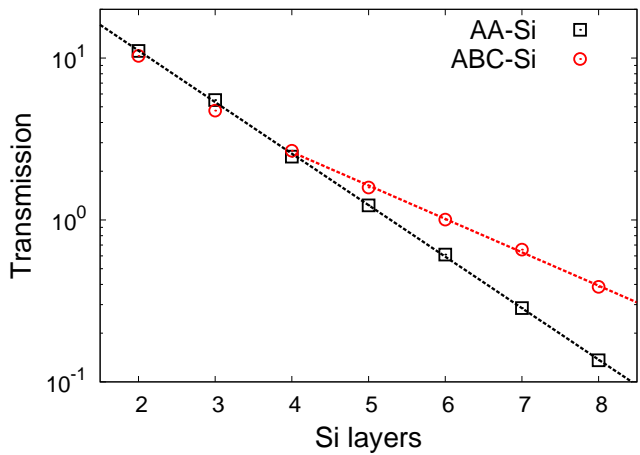}
\caption{
\label{fig:trans}  
(Color online)
The averaged transmission over ${\bf k}_{\perp}$
for Ag$|$AA-Si$|$Ag and Ag$|$ABC-Si$|$Ag junctions as a function of 
the number of Si layers.
}
\end{figure}

The transmission of Ag(111)$|$multilayer-silicene$|$Ag(111) junctions at the Fermi energy were calculated 
using the DFT-NEGF method, averaged over ${\bf k}_\|$-points in the first Brillouin zone,
\begin{equation}
\label{eq:mavgT}
\mathcal { T } = 
\frac{1} { N_{{\bf k}_\|} }
\sum_{{\bf k}_\| } T({\bf k}_\|).
\end{equation}
Because there is more than one transverse mode for each ${\bf k}_\perp$ due to the band structure folding,
the averaged transmission $\mathcal{T}$ could be larger than unity.
For thinner junctions, with less than four silicene layers,
the transmissions of AA-Si junctions are very close to those of ABC-Si junctions.
For thicker barriers, the averaged transmission decays exponentially as a function of the number of silicene layers.
The decay rate of transmission of ABC-Si junctions is smaller than that of AA-Si junctions.

\begin{figure}[h]
\includegraphics[width=0.8\linewidth]{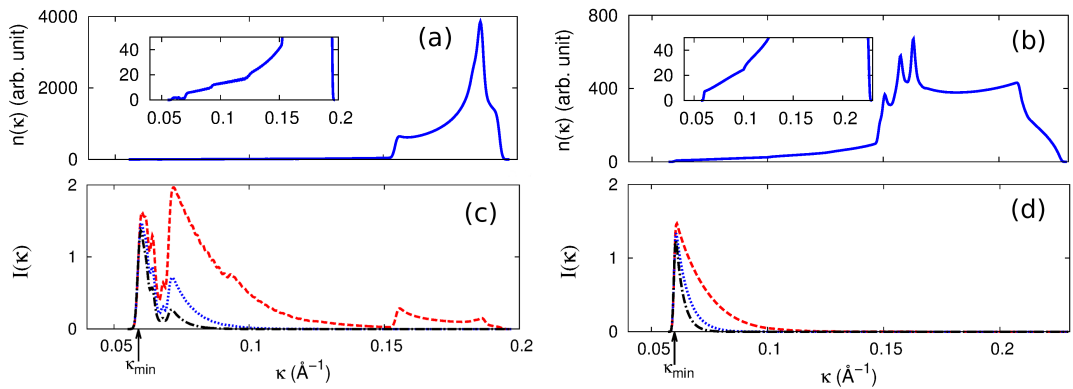}
\caption{
\label{fig:kappa-DOS}  
(Color online)
%{\color{blue}
The density of $\kappa$, denoted as $n(\kappa)$, for (a) AA-Si and (b) ABC-Si respectively;
$\kappa$ is defined as $\kappa({\bf k}_\|) = \min_{{\bf k}_\|} \Im (k_z)$; and
the relative importance of $\kappa \, I(\kappa) = n(\kappa)e^{-\kappa d}/[n(\kappa_{\mathrm{min}})e^{-\kappa_{\mathrm{min}} d}]$
for (c) AA-Si and (d) ABC-Si respectively, 
with $d=2$  (dashed, red lines), $ d=4 $ (dotted, blue lines), and $d =6 $ (dash-dot, black lines) 
times of the Si interlayer distance ($3.15\,\Ang$).
%}
}
\end{figure}

%{\color{blue} %% 2015 04 24
The tunneling through metal$|$semiconductor$|$metal junctions
can be understood in terms of the complex band structure of the semiconductor barrier.\cite{PhysRevLett.85.1088}
The tunneling current is carried out by evanescent states near the metal$|$semiconductor interfaces on the semiconductor side.
The energy dispersions of the evanescent and propagating states are called the complex band structure.
The junction is assumed to be periodic in the $x$-$y$ plane, 
so the in-plane components of the $k$-vector ${\bf k}_\|=(k_x,k_y)$ are always real,
while the component along the $z$-direction $k_z$ is allowed to be complex.
The Fermi energy of the junction lies in the band gap region of the semiconductor barrier,
so the imaginary part of $k_z$ is always nonzero.
Although there are infinitely many $k_z$ for each ${\bf k}_\|$,
we only considered the one with the smallest imaginary part (denoted as $\kappa$ hereafter),
since the corresponding evanescent state has the slowest decay rate.

We calculated $\kappa$ for each ${\bf k}_\|$,
where all of the ${\bf k}_\|$'s form a uniform mesh in the two-dimensional first Brillouin zone.
Following Ref.~\onlinecite{PhysRevLett.85.1088}, the density of $\kappa$ is defined as
$n(\kappa') = \sum_{{\bf k}_\|} \delta[\kappa({\bf k}_\|)-\kappa']$.
In practice the $\delta$-function is replaced by a Gaussian function.
The calculated density of $\kappa$ for AA-Si and ABC-Si are plotted in Fig.~\ref{fig:kappa-DOS}(a) and (b), respectively.
The density of $\kappa$ in the high-$\kappa$ region ($\kappa > 0.15/\Ang$)
is much larger than that in the low-$\kappa$ region ($\kappa < 0.15/\Ang$).

Not all of the $\kappa$'s are important for the tunneling.
One $\kappa$ is important if the number of $\kappa$ is large,
and/or if the corresponding transmission probability $e^{-2\kappa d}$ is large; 
the relative importance of each $\kappa$ is defined as
$I(\kappa) = n(\kappa)e^{-\kappa d}/[n(\kappa_{\mathrm{min}})e^{-\kappa_{\mathrm{min}} d}]$,
where $d$ is the thickness of the semiconductor barrier.\cite{PhysRevLett.85.1088}
The relative importance of the smallest $\kappa$ is scaled to be 1.
For very large $d$ the relative importance of all $\kappa$ except $\kappa_{\mathrm{min}}$ is zero, 
since $I(\kappa) \propto e^{-2(\kappa-\kappa_{\mathrm{min}})d}$,
and the tunneling is only contributed by the evanescent state corresponding to $\kappa_{\mathrm{min}}$, 
but for thin barriers, $\kappa$'s larger than $\kappa_{\mathrm{min}}$ can also be important.

The calculated relative importance curves for several different thicknesses of the silicon barrier
with  AA-Si and ABC-Si stacking orders are shown in Fig.~\ref{fig:kappa-DOS}(c) and (d) respectively,
for the thicknesses $d$ of Si barriers are corresponding to 2, 4, and 6 atomic layers.
The thickness $d$ is estimated to be smaller by 2 atomic layers than the actual thickness of the Si barrier,
since the interface-induced changes in the potential are confined in the first Si atomic layer at the interface.
Thus Figs.~\ref{fig:kappa-DOS}(c,d) are corresponding to junctions with 4, 6, and 8 Si atomic layers.
The range of important $\kappa$'s, i.e., the $\kappa$'s with a significant $I(\kappa)$, is larger for AA-Si than ABC-Si,
which explains the faster decay of the transmission of AA-Si based junctions.
%} %% 2015 04 24

%%%%%%%%%%%%%%%%%%%%%%%%%%%%%%%%%%%%%%%%%%%%%%%%%%%
\subsection{Resistance of junctions}
%%%%%%%%%%%%%%%%%%%%%%%%%%%%%%%%%%%%%%%%%%%%%%%%%%%

\begin{figure}[h]
\includegraphics[width=0.5\linewidth]{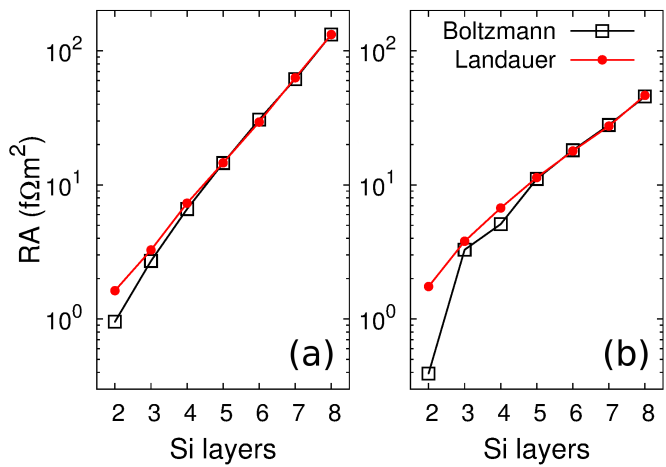}
\caption{
\label{fig:RA}  
(Color online)
Resistance-area products of (a) Ag$|$AA-Si$|$Ag and (b) Ag$|$ABC-Si$|$Ag junctions as a function 
of the number of Si layers calculated using the SCB equation (black rectangles) 
converge to those calculated using the Landauer formula (red dots).
}
\end{figure}

Using the Landauer formula, the resistance-area products (RA) of these junctions is
\begin{equation}
\label{eq:LandauerRA}
\mathrm{RA} = \frac{1}{\mathcal{T}} \frac{2 \pi \hbar }{2e^2} A,
\end{equation}
where $\mathcal{T}$ is the averaged transmission defined in Eq.~\eqref{eq:mavgT},
$2\pi\hbar/2e^2=12.9\,\mathrm{k\Omega}$ is the quantum resistance,
and $A$ is the cross section of the junction unit cell in the $x$-$y$ plane.
The resistance calculated using the Landauer formula Eq.~\eqref{eq:LandauerRA}
corresponds to the resistance measured using the ``two-probe'' configuration.
The two-probe resistance can be interpreted as the corresponding four-probe resistance in series with a contact resistance.
The contact resistance is the result of the mismatch in the number of channels in the leads and in the reservoirs:
there are a finite number of channels in the leads, but infinite in the reservoirs.\cite{Datta1995}
The two-dimensional multilayer-silicene junctions in question actually have infinite channels in their leads.
As a result, the Landauer formula Eq.~(\ref{eq:LandauerRA}) becomes invalid in the large-transmission region.

%{\color{blue}
To resolve this problem,
%}
we then used the SCB theory to calculate the RA of junctions,
which correspondis to the resistance measured in the ``four-probe'' configuration.
The values of RA calculated using the Landauer formula and the SCB theory are presented in Fig.~\ref{fig:RA}.
The SCB theory gives almost the same RA as the Landauer formula for junctions with more than five layers of silicene,
where the averaged transmission per channel is about $ 0.05 $.
In this small transmission limit, the contact resistance is negligible with respect to the junction resistance;
thus the two-probe resistance is nearly equal to the four-probe resistance.
This is the first time that the SCB theory is applied down to the small transmission region.
%{\color{blue}
We note that in this small transmission region, the Boltzmann equation becomes
numerically difficult to solve;  
as a result, it is more convenient to use the Landauer formula instead.
%}

The ABC-Si junction with two Si atomic layers has a RA $50\%$ smaller than that of the corresponding AA-Si junction,
indicating that the calculated resistance using the SCB theory is not entirely determined by the averaged transmission.
The existing various multi-channel extensions\cite{PhysRevB.31.6207,DiVentra} of the four-probe Landauer formula
have the common feature that the group velocities of Bloch states in the leads play a role.
The factors determining the SCB resistance are however rather difficult to analyze, 
due to the self-consistent nature of the Boltzmann equation.
The most important observation from Fig.~\ref{fig:RA} is that the upper limit of the transmission per channel 
for the applicability of the Landauer formula is equal to $ 0.05 $.
In the next section, we studied  \hbn-based junctions to explore whether this value ($ 0.05 $) is universal.

%%%%%%%%%%%%%%%%%%%%%%%%%%%%%%%%%%%%%%%%%%%%%%%%%%
\section{multilayer hexagonal Boron Nitride junctions}
%%%%%%%%%%%%%%%%%%%%%%%%%%%%%%%%%%%%%%%%%%%%%%%%%%
\label{sec:hbn}

In this section, we turn to junctions based on the wide-gap insulator hexagonal boron nitride (\hbn).
Hexagonal boron nitride is generally recognized as a good insulator because of its large energy band gap.
However, first-principles calculations in the literature\cite{PhysRevB.80.035408,PhysRevB.84.153406}
showed that the transmission of monolayer \hbn-based junctions is on the order of unity.
According to the preceding results on multilayer silicene junctions,
the Landauer formula is no longer applicable, but instead 
 SCB theory is needed to calculate the resistivity of monolayer \hbn{} junctions.

\begin{figure}[h]
\includegraphics[width=0.5\linewidth]{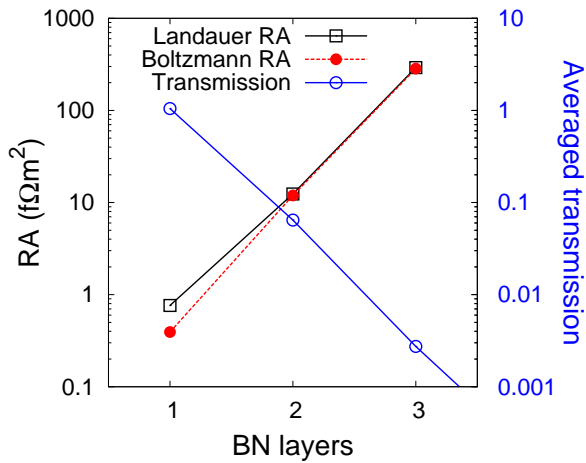}
\caption{
\label{fig:BN-RA}  
(Color online)
Resistance-area products of Ni(111)$|$\hbn$|$Ni junctions with different numbers of \hbn{} layers
calculated using the Landauer formula (black rectangles) and using the SCB theory (red dots).
The averaged transmissions are denoted by blue circles.
}
\end{figure}

The electrode for \hbn{} junctions is chosen to be {\it fcc}-Ni because of the small lattice mismatch,  
$2.504\,\Ang$ for \hbn{} versus $2.492\,\Ang$ for the (111)-surface of Ni.
The in-plane lattice constant of \hbn{} is squeezed slightly to match that of Ni(111), 
in order to simulate the commensurate \hbn/Ni(111) interface.
The distance between \hbn{} layers is set to the interlayer distance in bulk \hbn{}, $ 3.33\,\Ang$.
At the \hbn/Ni(111) interface, N atoms sit on the top of surface Ni atoms,
and the distance between \hbn{} and Ni(111) surfaces is $ 2.1\,\Ang $.\cite{PhysRevLett.79.4609}

The transmissions of mono-, bi-, and tri-layer \hbn{} based junctions
are calculated using the same method for Ag$|$silicene$|$Ag junctions,
except that spin-polarized calculations were carried out; 
the magnetic moments in the two Ni(111) leads are set to be parallel to each other.
The calculated transmission decays exponentially as a function of number of  \hbn{} layers (Fig.~\ref{fig:BN-RA}),
in accord with previous calculations.\cite{PhysRevB.84.153406}
No spin flip is considered in calculating the resistance, 
and the two spin channels are considered to be independent.
The RA of monolayer \hbn-based junctions calculated using the Landauer formula (Eq.~\eqref{eq:LandauerRA})
is about two times larger than that using the SCB theory, as shown in Fig.~\ref{fig:BN-RA}.
The difference in RA between Landauer formula and the SCB theory becomes very small
for junctions with more than one \hbn{} layer.
The junction with bilayer \hbn, 
which is the threshold for the Landauer formula to be applicable.
has a transmission of about $ 0.06 $.
We note that this value is very close to the value 0.05 of the critical transmission for multilayer-silicene junctions.

%%%%%%%%%%%%%%%%%%%%%%%%%%%%%%%%%%%%%%%%%%%%%%%%%%
\section{Summary}
%%%%%%%%%%%%%%%%%%%%%%%%%%%%%%%%%%%%%%%%%%%%%%%%%%
\label{sec:summary}

In this work we extended our previous work\cite{PhysRevB.88.125428} to multilayer silicene junctions with up to eight layers, 
and also compared with \hbn{} junctions.
Two stacking orders AA and ABC of multilayer-silicene were considered.
The calculated transmission decays as a function of the barrier thickness for junctions with more than four silicene layers.
The electrical resistances were calculated using the Landauer formula and the SCB theory.
We observed that the SCB resistance is not entirely determined by the averaged transmission.

Most importantly, we learned from calculations on multilayer-silicene junctions that
the upper limit of the transmission per channel for the applicability of the Landauer formula is $ 0.05 $,
above which the Landauer formula significantly overestimates the SCB resistance.
Additional calculations on the \hbn{} junctions also give a very close value ($ 0.06 $) for the critical transmission.

The SCB theory to calculate the four-probe resistance of junctions is revisited.
We focused on the CPP configuration, in particular we present the numerical method to solve the Boltzmann equation in details.
We believe that this work and our former work (Ref.~\onlinecite{PhysRevB.88.125428})
boost the applications of the SCB theory.

\begin{acknowledgments}

This work was supported by the US Department of Energy (DOE),
Office of Basic Energy Sciences (BES),
under Contract No. DE-FG02-02ER45995.
This research used resources of the National Energy Research Scientific Computing (NERSC) Center.

\end{acknowledgments}

%%%%%%%%%%%%%%%%%%%%%%%%%%%%%%%%%%%%%%%%%%%%%%%%%%%%%%%%%%%%%%%%%%%%%%%%%%%%%%%%%%%%%%%%%%%%%%%%%%%%%%%%%%%%%%%%%%%%%%%%%%%%%%%
%%%%%%%%%%%%%%%%%%%%%%%%%%%%%%%%%%%%%%%%%%%%%%%%%%%%%%%%%%%%%%%%%%%%%%%%%%%%%%%%%%%%%%%%%%%%%%%%%%%%%%%%%%%%%%%%%%%%%%%%%%%%%%%
%\newpage

\appendix

%%%%%%%%%%%%%%%%%%%%%%%%%%%%%%%%%%%%%%%%%%%%%%%%%%%%%%%%
\section{Numerical solution for the Boltzmann equation}
%%%%%%%%%%%%%%%%%%%%%%%%%%%%%%%%%%%%%%%%%%%%%%%%%%%%%%%%
%\label{sec:cpp}

In the appendix the numerical method for solving the Boltzmann equation is presented.
The Boltzmann equation for the CPP geometry is,
\begin{equation}
\label{eq:aboltzcpp}
\left[
    v_{z}^j({\bf k}_{\|}) \frac{\partial}{\partial z}
    + \frac{1}{\tau}
\right]
 h^j(z,{\bf k}_{\|})
-\frac{\mu(z)}{\tau}
=
-e v^j_{z}({\bf k}_{\|})
{\cal E}_z.
\end{equation}
%{\color{blue}
The solution of Eq.~\eqref{eq:aboltzcpp} can be written as the sum $h^j = u^j + w^j$.
The first term $u^j$ satisfies Eq.~\eqref{eq:aboltzcpp} for $\mu(z)=0$,
%}
\begin{equation}
\label{eq:aequ}
\left[
    v_{z}^j({\bf k}_{\|}) \frac{\partial}{\partial z}
    + \frac{1}{\tau}
\right]
 u^j(z,{\bf k}_{\|})
=
-e v^j_{z}({\bf k}_{\|})
{\cal E}_z ,
\end{equation}
for which we can write an analytical solution for $u^j$, 
\begin{equation}
\label{eq:au}
u^j(z,{\bf k}_{\|})
=-ev^j_z({\bf k}_{\|}){{\cal E}_z}\tau
+A^j({\bf k}_{\|})\exp\left[-\frac{z}{v^j_z({\bf k}_{\|})\tau}\right] ; 
\end{equation}
the $A^j$ are parameters to be determined by the boundary conditions.
Then, $w^j$ obeys 
\begin{equation}
\label{eq:aeqw}
v^j_{z}({\bf k}_{\|})\frac{\partial w^j(z,{\bf k}_{\|})}{\partial z} +
\frac{w^j(z,{\bf k}_{\|})-\mu(z)}{\tau} 
= 0. 
\end{equation}
Here, approximating the $z$ dependence of $\mu(z)$ to second order,
\begin{equation}
\label{eq:amuz}
\mu(z)=\mu_0+\mu_1z+\frac{1}{2}\mu_2z^2,
\end{equation}
we find the solution for $w^j$ is,
\begin{equation}
\label{eq:aw}
w^j(z+\Delta z,{\bf k}_{\|})=w^j(z,{\bf k}_{\|})
\exp\left[
       -\frac{\Delta z}{v^j_z({\bf k}_{\|})\tau}
\right]
-b_0\mu_0- b_1\mu_1 \Delta z-\frac{1}{2}\, b_2\mu_2 \Delta z^2,
\end{equation}
with
\begin{eqnarray}
\label{eq:ab0}
b_0&=& \exp\left[-\frac{\Delta z}{v^j_z({\bf k}_{\|})\tau}\right]-1, \\
\label{eq:aa22}
b_1 &=& -\frac{v^j_z({\bf k}_{\|})\tau}{\Delta z}b_0-1, \\
\label{eq:ab2}
b_2 &=& -2\frac{v^j_z({\bf k}_{\|})\tau}{\Delta z} b_1 -1.
\end{eqnarray}
The $z$-axis is uniformly discretized in practical calculations.
The value of $w^j$ on a grid point can be derived from the value on one of its neighbors,
since Eq.~\eqref{eq:aw} is a recursive type equation.

The boundary conditions for the distribution functions in leads are determined by
the transmission and reflection coefficients of the junctions between leads.
Here we only considered the systems consisting of one single junction sandwiched by two leads;
the extension to multiply leads and junctions is straightforward.
Suppose the junction between the leads is located at $z=z_0$.
The leads above ($z>z_0$) and below ($z<z_0$) the junction are denoted as the ``right'' ($R$) and the ``left'' ($L$) leads, respectively.
The boundary conditions for the distribution function are,
\begin{eqnarray}
\label{eq:abc1}
h_{>}^{R,j}(z_0^+) &=& 
\sum_{j'}^{N_R} |r^{RR}_{j,j'}|^2 \, h_{<}^{R,j'}(z_0^+)
+  \sum_{j'}^{N_L} |t^{LR}_{j,j'}|^2 \, h_<^{L,j'}(z_0^-) \\
\label{eq:abc2}
h_{>}^{L,j}(z_0^-) &=& 
\sum_{j'}^{N_L} |r^{LL}_{j,j'}|^2 \, h_{<}^{L,j'}(z_0^-)
+  \sum_{j'}^{N_R} |t^{RL}_{j,j'}|^2 \, h_<^{R,j'}(z_0^+)
\end{eqnarray}
The ${\bf k}_\|$ dependence of the distribution function $h$ and the transmission and reflection coefficients ($t$ and $r$) are omitted in Eqs.~(\ref{eq:abc1},\,\ref{eq:abc2}).
The total number of channels in the left and the right leads are $2N_L$ and $2N_R$ respectively,
half of them propagating towards ($<$) the junction and the other half against ($>$) the junction.

There is indeed a freedom to choose the boundary conditions for $v^j$ and $w^j$,
and  in our implementation both $v^j$ and $w^j$ satisfy the boundary conditions Eqs.~(\ref{eq:abc1},\,\ref{eq:abc2}).
Note that $u^j$ in Eq.~\eqref{eq:au} is decoupled for different ${\bf k}_{\|}$'s,
so do the boundary conditions in Eqs.~(\ref{eq:abc1},\,\ref{eq:abc2}).
At each ${\bf k}_{\|}$, there are $2N_L$ and $2N_R$ unknowns $A^j$'s in the left and right leads respectively,
so the total number of unknowns $A^j$ is $2N_L + 2N_R$.
The boundary conditions in Eqs.~(\ref{eq:abc1},\,\ref{eq:abc2}) provide $(N_L + N_R)$ equations,
which is less than the number of unknowns by $(N_L+N_R)$.
In fact, we considered that the leads have finite lengths, {\it i.e.},
the left lead extends down to $z=z_L$ and the right lead up to $z=z_R$ ($z_R > z_0 > z_L$).
The boundaries at $z=z_L$ and $z_R$ provide boundary conditions similar to Eqs.~(\ref{eq:abc1},\,\ref{eq:abc2}) but without the transmission part;
thus they provide another $N_L + N_R$ equations for determining the unknowns $A^j$.
The reflection coefficients of the boundaries at $z=z_L$ and $z_R$ are arbitrary
and they do not change the resistance of the junction if the length of both leads are long enough.

The $w^j$ in Eq.~\eqref{eq:aeqw} is no longer decoupled for different ${\bf k}_{\|}$,
because the chemical potential $\mu$ contains a summation over all the ${\bf k}_{\|}$'s.
As a result, Eq.~\eqref{eq:aeqw} for $w^j$ is a self-consistent field equation,
which can be solved using an iterative algorithm: 
given an initial guess for the chemical potential $\mu_{\mathrm{in}}(z)$,
$w^j$ can be solved using the same method to solve $u^j$.
After solving for $w^j(z)$, a new chemical potential $\mu_{\mathrm{out}}(z)$ can be built and used for the next iteration step.
The iteration stops when $\mu_{\mathrm{in}}(z)$ agrees with $\mu_{\mathrm{out}}(z)$ within some predefined numerical accuracy.
Efficient mixing algorithms such as Ref.~\onlinecite{PhysRevB.38.12807} can be used to accelerate the convergence.
The typical number of iteration steps requried for current-density conservation is several hundreds.

%{\color{blue}
%%%%%%%%%%%%%%%%%%%%%%%%%%%%%%%%%%%%%%%%%%%%%%%
\section{The expression for the local chemical potential}
%%%%%%%%%%%%%%%%%%%%%%%%%%%%%%%%%%%%%%%%%%%%%%%

We  derive in this Appendix the expression for the local chemical potential $\mu(z)$.
The number of electrons at position $z$ with energy $E$ is $\sum_{j, {\bf k}_\|}  f^j(z,{\bf k}_\|,E)$.
The electrons at $z$ are in local equilibrium with a local chemical potential $\mu({\bf r})$, {\it i.e.},
the occupation number of each mode is equal to $1/(e^{\beta [E-\mu(z)]}+1)$,
where $\beta=1/k_BT$, so we have
\begin{equation}
\sum_{j, {\bf k}_\|}  f^j(z,{\bf k}_\|,E) = \sum_{j, {\bf k}_\|} \frac{1}{e^{\beta [E-\mu(z)]}+1},
\end{equation}
or, averaged over $j$ and ${\bf k}_\|$, 
\begin{equation}
\label{eq:B_f}
\left\langle f^j(z,{\bf k}_\|,E) \right\rangle_{j, {\bf k}_\|}
=  \frac{1}{e^{\beta [E-\mu(z)]}+1}
\equiv {\cal F}[E-\mu(z)] .
\end{equation}
The equilibrium distribution function $f_0$ denotes  
electrons that are in global equilibrium with a global chemical potential $\mu_0$, 
\begin{equation}
\label{eq:B_f0}
f_0(E) = {\cal F}(E-\mu_0).
\end{equation}
From Eqs.~(\ref{eq:B_f}, \ref{eq:B_f0}) in Eq.~\eqref{eq:define_h}, we have,
\begin{equation}
-\left\langle h^j(z,{\bf k}_\|) \right\rangle_{j, {\bf k}_\|} \frac{d f_0 (E)}{d E}
={\cal F}[E-\mu(z)] - {\cal F}(E-\mu_0).
\end{equation}
In the linear-response regime, the value of $\mu(z)$ remains close to $\mu_0$, and 
\begin{equation}
{\cal F}[E-\mu(z)] - {\cal F}[E-\mu_0] =
\left . - [ \mu(z) - \mu_0 ] \, \frac{d {{\cal F}(x)}}{dx} \right |_{x=E-\mu_0}  .
\end{equation}
At low temperatures, we have $-df_0(E)/dE=-\left . {d {{\cal F}(x)}}/{dx} \right |_{x=E-\mu_0} = \delta(E-\mu_0)$.
As a result, the expression for the local chemical potential is 
\begin{equation}
\label{eq:ahmu}
\mu(z) = \left\langle h^j(z,{\bf k}_\|) \right\rangle_{ j, {\bf k}_\| } + \mu_0.
\end{equation}
Suppose that $h^j(z,{\bf k}_\|)$ is the solution of the Boltzmann equation with $\mu(z)$.
If the local chemical potential $\mu(z)$ is shifted by a constant $C$,
the corresponding solution of the Boltzmann equation becomes $h^j(z,{\bf k}_\|)+C$.
Both the voltage drop and the current density are independent of the constant $C$ 
according to their definitions.
So, we drop the term $\mu_0$ in the expression of $\mu(z)$ and obtain finally 
\begin{equation}
\label{eq:ahmu}
\mu(z) = \left\langle h^j(z,{\bf k}_\|) \right\rangle_{ j, {\bf k}_\| }.
\end{equation}

%} %% blue
%%%%%%%%%%%%%%%%%%%%%%%%%%%%%%%%%%%%%%

%\bibliography{../../bib-collection}

%%%%%%%%%%%%%%%%%%%%%%%%%%%%%%%%%%%%%%


\end{document}